\documentclass[twocolumn,superscriptaddress,amsmath,amssymb,showpacs,prb]{revtex4-1}

\usepackage{graphicx}
\usepackage{color}
\renewcommand{\phi}{\varphi}

\begin{document}

\title{Spatial decomposition of magnetic anisotropy in magnets: application for doped $\text{Fe}_{16}\text{N}_{2}$}

\author{Yang Sun}
    \affiliation{Department of Applied Physics and Applied Mathematics, Columbia University, New York, NY, 10027, USA}
\author{Yong-Xin Yao}
    \affiliation{Ames Laboratory, US DOE and Department of Physics, Iowa State University, Ames, Iowa 50011, USA}
\author{Manh Cuong Nguyen}
    \affiliation{Ames Laboratory, US DOE and Department of Physics, Iowa State University, Ames, Iowa 50011, USA}
\author{Cai-Zhuang Wang}
    \affiliation{Ames Laboratory, US DOE and Department of Physics, Iowa State University, Ames, Iowa 50011, USA}
\author{Kai-Ming Ho}
    \affiliation{Ames Laboratory, US DOE and Department of Physics, Iowa State University, Ames, Iowa 50011, USA}
\author{Vladimir Antropov}
    \email[Email: ]{antropov@ameslab.gov}
    \affiliation{Ames Laboratory, US DOE and Department of Physics, Iowa State University, Ames, Iowa 50011, USA}

\date{Oct. 23, 2020}

\begin{abstract}

We propose a scheme of decomposition of the total relativistic energy in solids to intra- and interatomic contributions.  The method is based on a site variation of such fundamental constant as the speed of light. As a practical illustration of the method, we tested such decomposition in the case of a spin-orbit interaction variation for the decomposition of the magnetic anisotropy energy (MAE) in CoPt. We further studied the $\alpha''-\text{Fe}_{16}\text{N}_{2}$ magnet doped by Bi, Sb, Co and Pt atoms. It was found that the addition of Pt atoms can enhance the MAE by as large as five times while Bi and Sb substitutions double the total MAE. Using the proposed technique, we demonstrate the spatial distribution of these enhancements. Our studies also suggest that Sb, Pt and Co substitutions could be synthesized by experiments.

\end{abstract}

\maketitle

\section{Introduction}
Today, studies of magnetic anisotropy (MA) are very popular due to many existing and potential applications of certain magnets as well as a very rich microscopic physics of these materials. The strength of MA effects is usually relatively small, and because of this they have been called for years as “secondary” magnetic effects. Initial MA models \cite{1,2,3} employed very simplified ideas of “single ion” anisotropy and could not be applied for metallic magnets, which represent by far the most powerful magnets currently. MA in metals has a very rich and complicated physics as a function of concentration and temperature (see, for instance, “magnetic chameleon” in Refs. \cite{4,5}) which requires detailed knowledge of the electronic structure at and around the Fermi level. Theoretically, very extended studies of MA phenomena have been done based on the traditional k-space analysis \cite{6,7}. For many theoretical models, however, a physical picture based on real space decomposition of relativistic magnetic interactions can often be more useful. From the point of view of searching for new magnetic materials, the replacement of some atoms (chemical doping) is a very common procedure (see a recent review \cite{8}). Due to all these factors, questions such as: “how long-ranged are anisotropic interactions?”, “when can we use a single-ion approximation?” and “what is the influence of hybridization on the spatial dependence of magnetic anisotropy?” require answers in specific materials, especially metals, where all electronic interactions are expected to be long-ranged due to Fermi surface effects. Thus, it is needed to be able to decompose the observed relativistic properties in the crystals to intra and interatomic contributions including possible multiatomic interactions.

Such decomposition to on-site and pairwise interactions can be done using spin-orbit (SO) coupling as a perturbation with a Green function formalism \cite{9,10}. However, such methods are difficult to implement in modern full potential electronic structure studies that are based mostly on the non-orthogonal basis set Hamiltonian constructions. This is unfortunate as the majority of high anisotropy systems must have non-cubic symmetry (mostly tetragonal or hexagonal) with clear importance of non-spherical terms in the potential. From a different perspective, the relativistic perturbation theory can be questionable in the materials where, for instance, crystal field effects are small relative to relativistic effects (rare earth atoms). Also, the traditional replacement of the total energy (TE) by its one -electron contribution \cite{9,10} is a very uncontrolled approximation and the opportunity to decompose the exact TE is really needed. Thus, a general method for the analysis of atomic and interatomic relativistic interactions in arbitrary systems in popular Hamiltonian based band structure methods is highly desired.

This paper presents a simple technique to solve this problem: we propose to consider the speed of light ($c$) in the description of certain valence electronic states of atom in the crystal as a variable. We first consider the spin-polarized Dirac equation. Inside atom our zero-order Hamiltonian includes the scalar potential $V(\textbf{r})$ and the effective magnetic field $\boldsymbol{B}(\boldsymbol{r})$:
\begin{equation}
H=c\boldsymbol{\alpha}\cdot\boldsymbol{p}+(\beta+1)mc^2+V(\boldsymbol{r})+\boldsymbol{\beta}\boldsymbol{\sigma}\cdot\boldsymbol{B}(\boldsymbol{r}) \label{eq1}.
\end{equation}

Here, $c$ is the speed of light,  $\boldsymbol{\alpha}$ and $\boldsymbol{\beta}$ are standard Dirac 4x4 matrices and $\boldsymbol{p}$ is the momentum operator. The three Cartesian components of the vector $\boldsymbol{\sigma}$  are the Pauli $2\times 2$ matrices, and $m$ is the electron rest mass. In the atomic Rydberg units $m=1/2$ and $c \approx 274$. Thus, formally one can vary the inverse of the speed of light from 0 (its value in the non-relativistic limit) to $\approx 1/274$ on all atomic sites or some of them. For each particular $c$ value, the TE of relativistic DFT can be obtained variationally without needing constraining fields. The resulting TE $E\{c_{i\alpha}\}$ would be a functional of site index $i$ and electronic quantum number (if needed). So, we assume that the relativistic TE can be presented as a functional of all $c_i$
\begin{equation}
E=E\{c_i\} \label{eq2}.
\end{equation}

Now we consider a variation of the TE $\delta E$  by changing the speed of light $c$ on a single atomic site $1$ while solving the atomic Dirac equation. Let us first assume nearly non-relativistic case ($(1/c_1)^2 = \alpha_1 \ll 1$). After solving the band problem the corresponding variation of the TE Eqn.\ref{eq2} can be presented as
\begin{equation}
\delta E = \alpha_1 N + \alpha_1^2 A + ... \label{eq3},
\end{equation}
where the first term corresponds to the linear variation of the TE, the second one to quadratic and so on. Here $A$ term represents the TE variation on site $1$ (intra-atomic contribution) while term with $N$ corresponds to energy change due to inter-atomic coupling of atom $1$ with its surrounding.  In a case of simultaneous variation of $c$ on two sites $1$ and $2$ the TE variation can be presented in a similar way as
\begin{equation}
\delta E = \alpha_1^2 A + \alpha_1 \alpha_2 C + \alpha_2^2 B + \alpha_1 N + \alpha_2 M... \label{eq4},
\end{equation}
where $A$ and $B$ are pure intra-atomic parts, $C$ term corresponds to the inter-atomic interaction of atom $1$ and $2$, and $N(M)$ terms describe pairwise inter-atomic couplings of atom 1 (2) with all other atoms (except atom 2(1)).

For the case of MA symmetric interactions, one can repeat describing the above action for different directions of the magnetization and for the MAE we will have similar decomposition
\begin{equation}
K = \alpha_1 N + \alpha_1^2 A + ... \label{eq5},
\end{equation}
with $A$ corresponding to the single site anisotropy contribution and $N$ to the sum of pairwise MA terms. The expansion \ref{eq5} already contains only small MA terms, while Eq. \ref{eq3} describes much larger changes of relativistic TE (including, for instance, the variation of scalar relativistic terms like Darwin and mass-velocity in a case of perturbation theory usage).

The expansions above can be also illustrated using a perturbation theory approach if we use the atomic SO coupling as a smallness parameter. Then the analog of Eq. \ref{eq4} for two sites $i$ and $j$ can be presented as
\begin{equation}
K = \lambda_i^2 A + 2 \lambda_i\lambda_j C + \lambda_j^2 B \label{eq6},
\end{equation}
where $\lambda_i$ is a SO coupling parameter on site $i$. This simple example demonstrates a basic idea of this paper: decomposition of the total relativistic energy or its angular variation (anisotropic torque) to intra- and inter-atomic terms. Evidently, considering the full angular dependence of E in magnetic relativistic cases, one can obtain the desired spatial dependencies of the symmetric anisotropic energy terms. In a case of anisotropic magnetic interactions (like Dzyaloshinski-Moriya interaction), the speed of light change would affect the small angle between moments on different sites and directly provides an opportunity to estimate the strength of this interaction as well. 

The main advantage of this scheme is an opportunity to obtain relativistic inter-site interactions using highly precise electronic structure codes with no shape approximation for the potential. Second, this scheme can be used for the exact decomposition of the TE with or without any relativistic perturbation theory formalism. Naturally it can be used for any approximate description of the TE as well, including its one electron contribution (see Ref. \cite{9,10}) or just SO energy analysis (see below).  

So far, we discussed a decomposition of the TE without any approximation. While there are no concerns about such calculations, in some cases one can use simplified methods to obtain the TE. For instance, there are many approximate ways of presenting the TE as a sum of the atomic contributions. As usual, such “quasi atomic” partial contributions represent a sum of pure atomic energy and inter-atomic contributions. One of the most popular techniques, for instance, is the replacement of the TE by the consideration of its one electron contributions in DFT-based methods (see Ref. \cite{11,12}).

While our proposed scheme is introduced for the fully relativistic spin-polarized approach  \cite{13}, it is still more convenient to combine this proposal with a perturbative scheme to separate small anisotropic terms from large isotropic ones. A qualitative advantage of the relativistic perturbative theory is that the major relativistic interaction responsible for the magnetic anisotropy can be separated from all others in the second order over $c^{-2}$ as the SO coupling. A variation of the speed of light described above, in turn, can be replaced by a variation of SO coupling alone. Earlier, we implemented a variation of SO coupling to study the applicability of relativistic perturbation theory and a relativistic virial theorem \cite{14} without discussion of inter-atomic interactions. Similar studies with the removal SO coupling have been done in Ref. \cite{15} to understand MAE for adatoms and monolayers.

Below we will concentrate on the analysis of MAE. The most valuable information we would like to obtain is the knowledge of key contributor to the enhancement of MAE. As we will demonstrate below, this can be done by creating a site anisotropy diagram as a function of a particular atomic SO coupling (or any of its electronic orbital component).  It can be obtained using a partial one electron contribution analysis or by the calculation of atomic SO coupling energy ($E_{so}$) which is directly related to the total relativistic energy change when SO coupling is added \cite{16}. We will be using this perturbation theory SO energy calculations due to the good accuracy of perturbation theory in d-metals. The analysis of one electron contribution has been used since the 60s \cite{9,10} and represent a very uncontrolled approximation. The “local force theorem” \cite{11,12} cannot be justified in this case, as it is formulated for variations only, while in the calculation of MAE one has to use the finite differences.

In this work, we first test our approach for the well-known magnet CoPt which can be considered as a prototype system with a non-trivial site decomposition of MAE. We then consider the more complicated magnet Fe$_{16}$N$_2$ doped by Bi, Sb, Co and Pt atoms. Both structural stability and MAE will be studied. The reason to choose these dopants is that they have very different electronic structures: Pt has a localized $d$ state, while Bi and Sb show a spreading delocalized $p$ state. By decomposing the calculated MAE in terms of atomic SO coupling anisotropies, we show in detail how the SO coupling interaction of the dopants with different electronic structures affect the MAE of Fe$_{16}$N$_2$ compounds. 

\section{Methods}

First-principles calculations were carried out using the density functional theory (DFT) with spin polarization. The generalized gradient approximation (GGA) in the form of the PBE (Perdew, Burke, and Ernzerhof \cite{17}) implemented in the VASP code \cite{18} was used. Kinetic energy cutoff was set to 650 eV. The Monkhorst-Pack’s scheme \cite{19} was used for Brillouin zone sampling with a k-point grid resolution of $2\pi \times 0.033 \text{\AA}^{-1}$  during the structure optimization. The ionic relaxation was stopped when the force on each atom became smaller than 0.01 eV/ $\text{\AA}$. The MAE calculations were also performed with the VASP code. All symmetry operations were switched off completely when the SO coupling was turned on. A denser k-point grid ($2\pi \times 0.016 \text{\AA}^{-1}$) was used in the MAE calculations to achieve a better k-point convergence. To obtain the MAE, we first performed a fully self-consistent collinear calculation. Then we started from the charge density and performed one-shot calculations with SO coupling and different orientations of magnetization direction aligned along [001] and [100] to get the total energies  $E^{[001]}$ and $E^{[100]}$, respectively \cite{20}. The SO coupling can also be calculated with self-consistent calculation. The decomposition scheme works for both cases.

\section{Results and Discussion}
We first consider a well-known CoPt ferromagnet. This layered tetragonal system (AuCu structural type) shows a very large MAE (comparable to those in rare-earth magnets) and has been studied many times in the past. All calculational parameters have been taken as in Ref. \cite{14} with k-point mesh $24 \times 24 \times 26$. The unusual feature of this system is that the MAE in CoPt (as shown in \cite{16}) does not follow the anisotropy of the orbital moment and cannot be described by Streever’s model \cite{9,21}.

Our calculated total MAE is 0.84 meV and it agrees well with earlier full-potential results \cite{14,16} and the experiment \cite{22}. As we mentioned above, to obtain the desired spatial decomposition of this number, one can use the exact TE calculations as we suggested in the beginning and/or approximate methods like one-electron and SO coupling energies calculated at each site. Later two are closely related. Within the second-order perturbation theory, it has been shown \cite{16} that the total MAE of a hexagonal or tetragonal crystal can be written as
\begin{equation}
K_{\text{MAE}}=\Sigma_i K_{\text{SO}}(i)/2
\label{eq7},
\end{equation}
where $K_{\text{SO}}(i)$ is the anisotropy of SO coupling energy from the atom $i$ and the summation includes all atoms in the unit cell. $K_{\text{SO}}(i)=E_{\text{SO}}^{[100]}(i)-E_{\text{SO}}^{[001]}(i)$, where $E_{\text{SO}}^{[001]}(i)$ and $E_{\text{SO}}^{[100]}(i)$  are SO coupling energies of the atom $i$ with the magnetization aligned in the [001] and [100] directions, respectively. 

For simplicity, we consider the same change for all atomic valence electrons, but any orbital decomposition can be done in the same way. Technically, we introduce an artificial modification of SO coupling strength on the doping site to investigate how $K_{so}(i)$ on the other atoms changes by tuning the SO coupling strength on the dopant. Within the standard relativistic perturbation theory \cite{23}, the SO coupling Hamiltonian has the form
\begin{equation}
{\hat{H}}_{\text{SOC}}=\xi\hat{\sigma}\cdot\hat{L}
\label{hsoc},
\end{equation}
where $\hat{\sigma}$ is a Pauli spin operator, and $\hat{L}$ is the angular momentum operator. The SO coupling strength $\xi$ is proportional to the radial derivative of the spherical part of the effective all-electron potential $V(r)$ within the augmentation atomic sphere as 
\begin{equation}
\xi=\frac{\lambda}{c^2}\frac{\hbar^2}{{r(2m_e)}^2}\frac{dV(r)}{dr}
\label{xi}.
\end{equation}

Here we employ a scaling factor $\lambda$ in Eqn. ~\ref{xi} for any atom. We gradually vary $\lambda$ from 0 to 1 and in this case perform self-consistent calculations with SO coupling \cite{24} to obtain the corresponding TE and  $K_{so}(i)$ as the function of $\lambda$. In the case where one needs to extract any orbital dependent decomposition, a scaling factor $\frac{\lambda}{c^2}$ in SO Hamiltonian can be introduced just for a specific orbital.

Thus, we initially vary SO coupling on atoms inside a single layer $i$ in CoPt and calculate the resulting atomic MAE on each layer in our supercell. While the change of $K_{\text{SO}}(i)$ on-site $i$ will be a sum of the pure intra-atomic  (proportional to $\lambda_i^2$) and many intersite terms ($\lambda_i\lambda_j$), the change $K_{\text{SO}}(i)$ on other atoms will be the result of the intersite interactions only. Thus, one can clearly extract all pairwise interactions just from a variation of SO coupling on a given site. Functional dependence on $\lambda_i$ allows us to determine the effective spatial decomposition of anisotropy. 

\begin{figure}
\includegraphics[width=0.38\textwidth]{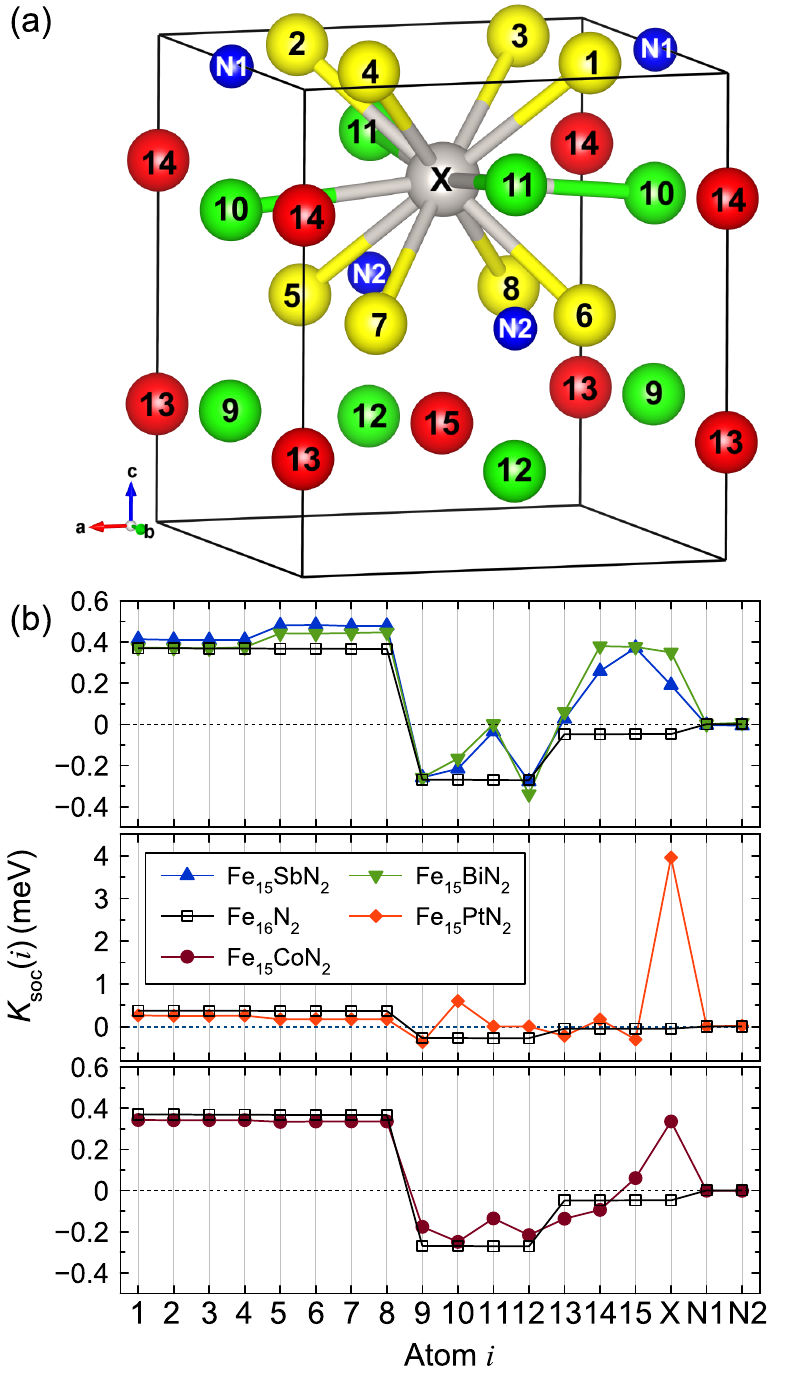}
\caption{\label{fig:fig1} (a) Crystal structure of $\text{Fe}_{16}\text{N}_{2}$ with doping X. Yellow are Fe-8h, red are Fe-4d, green are Fe-4h positions and blue are N atoms. X is the dopant at the 4d site. (b) The SO coupling anisotropy $K_{so}$ of each atom. Note the scale of the middle panel is different.}
\end{figure}

Our results indicate that the Pt atoms in CoPt produce the dominant part of the MAE (0.96 meV), while the Co subsystem has a small and negative (-0.085 meV) contribution to the total MAE. According to Eqn.\ref{eq1}, these quantities in turn can be presented as a corresponding sum of the pure on-site (``single-ion'') and all intersite terms. Using a variation of $\lambda$, we obtain the desired spatial decomposition. It appears that the pure on-site term dominates Pt atom MAE (0.85 meV) while the total contribution from the first nearest neighbor (NN) Co atoms ($K_{\text{Pt-Co}}$) is nearly 5-6 times smaller and negative (-0.15 meV). Simultaneously, the total pairwise contribution from the NN Pt atoms along the z-direction ($K_{\text{Pt-Co}}$) is positive and is the largest intersite interaction in the system (0.22 meV). Interlayer interactions beyond these two NN are smaller in amplitude and oscillating. The total MAE on Co atom is small and negative due to domination of negative pairwise contributions from Pt atoms (-0.15 meV). While Co atom on-site anisotropy is positive it is too small to compete even with negative pairwise Pt-Co interactions. Thus the Pt atom contributions are absolutely dominating providing a strong and positive Pt on-site anisotropy, somewhat smaller positive Pt-Pt and negative Pt-Co two-site anisotropies. These results are in qualitative agreement with those obtained a long time ago in Ref.\cite{10} using the Green function method and the atomic sphere approximation. Our results of anisotropy calculations using the TE and SO coupling analysis deviate from each other insignificantly (1-4 $\%$).

Next, we switch to the magnet $\text{Fe}_{16}\text{N}_{2}$. The pure $\text{Fe}_{16}\text{N}_{2}$ phase (space group I4/mmm) contains three Wyckoff positions for Fe: 4d, 4e and 8h. As shown in Fig.~\ref{fig:fig1}(a), we mainly consider the doping on the 4d position. The 4d position has 8 Fe-8h sites as the first NNs and 4 Fe-4e sites as the second NNs.

To obtain a reasonable estimate of energy stability, all the crystal structures are first relaxed by the DFT. The obtained lattice parameters are shown in Table \ref{table:tab1}. For pure $\text{Fe}_{16}\text{N}_{2}$, the volume from the DFT is slightly smaller than the room-temperature experimental data\cite{25} (a=5.72 $\text{\AA}$, c=6.29 $\text{\AA}$). This is reasonable as the calculation is done at $T=0K$. Doping with Bi, Sb and Pt atoms all expands the lattice. Doping with Co atom keeps the lattice almost unchanged. 

The energy stabilities are considered based on the formation energy ($E_f$) and the energy relative to the convex hull ($E_c$). Here we take $\text{Fe}_{15}\text{SbN}_{2}$ as an example to illustrate how the $E_f$ and $E_c$ are calculated. In this case, the formation energy is defined by
\begin{widetext}
\begin{equation}
E_f(\text{Fe}_{15}\text{SbN}_{2})=E(\text{Fe}_{15}\text{SbN}_{2})-\frac{15}{18}E(\text{Fe})-\frac{1}{18}E(\text{Sb})-\frac{2}{18}E(\text{N})
\label{ef},
\end{equation}

\end{widetext}
where $E(\cdot)$ is the total energy (per-atom) of the corresponding phase. In this case, the reference phases are the 0K ground-state structures of elementary Fe (bcc-$Im\bar{3}m$), Sb ($R\bar{3}m$) and N ($Pa\bar{3}$), respectively. A negative formation energy indicates the compound is energetically favorable against phase decomposition. The energy relative to the convex hull $E_c$ is relevant to the experimental synthesizability. The convex hull is composed of planes (curves if binary) connecting the formation energies of all thermodynamically stable phases. To compute the energy relative to the convex hull ($E_c$), one needs to consider the closest stable phases near the concertation, which can be found in the Material Project database \cite{26}. Here, for $\text{Fe}_{15}\text{SbN}_{2}$ , the surrounding phases on the ternary phase diagram are Fe ($Im\bar{3}m$), Fe3N ($P6_322$) and FeSb2 ($Pnnm$). Therefore, $E_c$ is calculated as
\begin{widetext}
\begin{equation}
E_c(\text{Fe}_{15}\text{SbN}_{2})=E(\text{Fe}_{15}\text{SbN}_{2})-\frac{8.5}{18}E(\text{Fe})-\frac{2}{18}E(\text{Fe}_3\text{N})-\frac{0.5}{18}E(\text{FeSb}_2)
\label{ef}.
\end{equation}
\end{widetext}

When $E_c=0$, the compound is thermodynamically stable. A compound with a smaller $E_c$ indicates a higher possibility to be synthesized in experiments. As a reference, it was found that 80$\%$ of compounds in the Inorganic Crystal Structure Database \cite{27} have an $E_c$ of less than 36 meV per atom \cite{28}.

Examining the energy values in Table \ref{table:tab1}, one can see that pure $\text{Fe}_{16}\text{N}_{2}$ has a very small $E_c$. This is consistent with the fact that it can be synthesized by experiment. It is also interesting to note that the compounds Sb, Pt and Co also have small values for $E_c$ (in a range of 20-30 meV/atom). This suggests that it is possible to form these compounds through experimental synthesis \cite{24}. In contrast, the Bi-doped $\text{Fe}_{16}\text{N}_{2}$ has positive formation energy and it is thermodynamically unstable against phase separation. Therefore, except for the Bi-doped $\text{Fe}_{16}\text{N}_{2}$, other doped compounds considered in this work would be achieved by experiments.

\begin{table}
\caption{\label{table:tab1} Optimized lattice parameters (a, c), the formation energy ($E_f$), the energy above the convex hull ($E_c$) and its reference.}
\centering
\begin{tabular} {  c | c | c | c | c | c  }
\hline
\hline
Compounds & a  & c  & $E_f$(meV & $E_c$(meV  & Reference phases \\
& ($\text{\AA}$) & ($\text{\AA}$) & /atom) & /atom) & in $E_c$ calculations \\
\hline
$\text{Fe}_{16}\text{N}_{2}$ & 5.68 & 6.22 & -33.3 & 4.5 & Fe, $\text{Fe}_{3}\text{N}$ \\
\hline
$\text{Fe}_{15}\text{BiN}_{2}$ & 5.76 & 6.34 & 63.2 & - & Fe, Bi, N \\
\hline
$\text{Fe}_{15}\text{SbN}_{2}$ & 5.72 & 6.30 & -8.8 & 31.0 & Fe, $\text{Fe}_{3}\text{N}$, $\text{FeSb}_{2}$ \\
\hline
$\text{Fe}_{15}\text{PtN}_{2}$ & 5.72 & 6.27 & -38.5 & 32.8 & Fe, $\text{Fe}_{3}\text{N}$, $\text{Fe}_{3}\text{PtN}$ \\
\hline
$\text{Fe}_{15}\text{CoN}_{2}$ & 5.68 & 6.21 & -27.8 & 20.7 & Fe, $\text{Fe}_{3}\text{N}$, $\text{Fe}_{3}\text{Co}$ \\
\hline
\hline
\end{tabular}
\end{table}

We now consider the effect of doping on the MAE in  $\text{Fe}_{16}\text{N}_{2}$. Table \ref{table:tab2} shows the MAE for the relaxed structures computed by VASP code \cite{18}  as described in the method section (denoted as $K_{\text{MAE}}$ with the unit of meV/cell and $K_1$ with the unit of MJ/$\text{m}^3$, respectively). Based on our calculations, the value of $K_1$ in pure $\text{Fe}_{16}\text{N}_{2}$ is 0.682 MJ/$\text{m}^3$. This is consistent with experimental reports\cite{29,30,31,32,33}  in a range of 0.44-2.0 MJ/$\text{m}^3$ as well as previous DFT calculations \cite{34,35,36}. Our calculations show that the MAE is improved by all considered dopants: compared to $\text{Fe}_{16}\text{N}_{2}$, Bi and Sb dopants increase the anisotropy by a factor of two, and Pt doping increases the MAE more than four times. Co doping provides a slight increase in MAE.

\begin{table}
\caption{\label{table:tab2} Magnetic anisotropy energy $K_{\text{MAE}}$, magnetocrystalline anisotropy constant ($K_1$), half of the total SO coupling anisotropy ($\sum K_{\text{SO}}/2$) and the difference between $K_{\text{MAE}}$ and $\sum K_{\text{SO}}/2$, i.e. $\Delta=\frac{\lvert K_{\text{MAE}}-\sum K_{\text{SO}}/2 \lvert }{K_{\text{MAE}}}$.}
\centering
\begin{tabular} {  c | c | c | c | c  }
\hline
\hline
Compounds & $K_{\text{MAE}}$  & $K_1$  & $\sum K_{\text{SO}}/2$ & $\Delta$\\
 & (meV/cell) & (MJ/$\text{m}^3$) & (meV/cell)  & ($\%$)  \\
\hline
$\text{Fe}_{16}\text{N}_{2}$ & 0.846 & 0.682 & 0.840 & 0.70 \\
\hline
$\text{Fe}_{15}\text{BiN}_{2}$ & 1.854 & 1.415	& 1.838 & 0.86 \\
\hline
$\text{Fe}_{15}\text{SbN}_{2}$ & 1.796 & 1.395 & 1.808 & 0.67 \\
\hline
$\text{Fe}_{15}\text{PtN}_{2}$ & 3.348 & 2.613 & 2.794 & 17.0 \\
\hline
$\text{Fe}_{15}\text{CoN}_{2}$ & 1.044 & 0.835 & 1.049 & 0.48 \\
\hline
\hline
\end{tabular}
\end{table}
\twocolumngrid

Next, we proceed with the decomposition of the total MAE into the contributions from different atomic sites to explore the spatial distribution of this MAE enhancement. In Table \ref{table:tab2}, we examine the validity of Eqn.\ref{eq7} in the current systems. It shows the relation is very valid for $\text{Fe}_{16}\text{N}_{2}$, $\text{Fe}_{15}\text{BiN}_{2}$, $\text{Fe}_{15}\text{SbN}_{2}$  and $\text{Fe}_{15}\text{CoN}_{2}$ (the deviations are all within 1\%). For $\text{Fe}_{15}\text{PtN}_{2}$, the discrepancy is much larger (17\%), indicating possible importance of higher order terms of perturbation theory.

\onecolumngrid

\begin{figure} [t]
\includegraphics[width=0.8\textwidth]{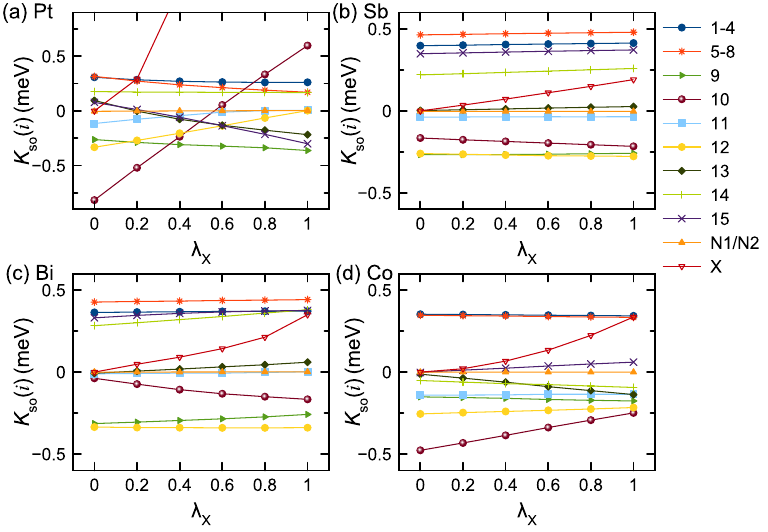}
\caption{\label{fig:fig2} The atomic SO coupling anisotropy energy $K_{\text{SO}}(i)$ as a function of the scaling factor $\lambda$ on the doping site (Fe16) for the four doped structures. For each curve all the $\lambda$ are fixed except the $\lambda_X$ varies.  Only 1, 5 and N1 are shown because 1-4, 5-8, N1/N2 sites are equivalent.}
\end{figure}
\twocolumngrid

In Fig.~\ref{fig:fig1} (b), the SO coupling anisotropy energies of each atomic site ($K_{\text{SO}}(i)$) in $\text{Fe}_{16}\text{N}_{2}$ and the four doped compounds are plotted. In pure $\text{Fe}_{16}\text{N}_{2}$, $K_{\text{SO}}$ of the 8h sites (sites Fe1-Fe8 in Fig. \ref{fig:fig1}a) promotes the MA along the easy axis (001) while 4h sites $K_{\text{SO}}$ (Fe9-Fe12) prefer an in-plane direction. $K_{\text{SO}}$ at the 4d sites (Fe13-Fe15 and X) overall are not significant. For the doped structure, the atoms are re-indexed because the symmetry of the original Wyckoff positions is broken. Compared to $\text{Fe}_{16}\text{N}_{2}$, the $K_{\text{SO}}$  for most Fe atoms in the unit cell is larger and positive upon Sb and Bi doping.  The most significant increases of $K_{\text{SO}}$  caused by Bi and Sb are at the sites Fe14 and Fe15. They are both initially 4d sites, and neither is the first NNs (Fe1-Fe8) or second NNs (Fe10 and Fe11) of the doped site. Therefore, the effect of doping on $K_{\text{SO}}$ \ is not very short-ranged but goes well beyond the nearest atomic shells of the dopants. The mechanism of change in $K_{\text{SO}}$ due to Pt doping is different from Sb and Bi doping. Pt induced a strong $K_{\text{SO}}$ by itself. While it slightly hindered the anisotropy along the [001] direction at 8h, Fe13 and Fe15 sites, the magnitude $K_{\text{SO}}$  of Pt is so huge that the overall MAE is still significantly enhanced. Co doping increased part of $K_{\text{SO}}$  along the easy axis (4h sites) but also increased the anisotropy of atoms along the hard axis (8d sites). This cancellation results in a very slight change of MAE from its pure $\text{Fe}_{16}\text{N}_{2}$ value.

To decompose the atomic magnetic anisotropies, we proceed with an adjustment of speed of light (or SO coupling of the valence electrons on a particular atomic site) described above.  Figure~\ref{fig:fig2} plots $K_{\text{SO}}$ on all sites in our supercell as a function of SO strength $\lambda_X$ on the doping site X. It shows that a change of $K_{\text{SO}}$  on all other atoms is a linear function of $\lambda_X$ because this change is induced by a pairwise interaction. In turn, dopant $K_{\text{SO}}$ changed as $\lambda_X^2$ and clearly demonstrated a quadratic behavior in Fig.~\ref{fig:fig2}. The $K_{\text{SO}}$ of Pt dopant is large (around 4 meV at $\lambda_X$=1) and is not shown in Figure~\ref{fig:fig2}(a). In all cases, the Fe10 site shows the strongest interaction with the dopant ($K_{\text{X-Fe10}}$ interaction) as its anisotropy $K_{\text{SO}}$ always changed most significantly with $\lambda_X$. In the case of Pt, the slope of anisotropy dependence on Fe12 site is also correlated with the slope of $K_{\text{SO}}$ on the Pt site (positive pairwise interaction). For other dopants, the anisotropies of majority atoms were almost independent on $\lambda_X$. This supported the somewhat ``single-ion'' character of dopant MA. On the other hand, different Fe sites exhibited weak but different sign changes of $K_{\text{SO}}$ as $\lambda_X$ changed from 0 to 1. This indicated different signs of pairwise interactions with this particular Fe-dopant. Figure ~\ref{fig:fig2} clearly shows that in many cases the pairwise interactions strongly influence some atomic anisotropies, changing both their amplitudes and signs. The first, second and even third NNs anisotropic interactions $K_{\text{Pt-Fe}}$ can be larger than pure single site contribution $K_{\text{Fe}}$. Overall, the MAE is not limited by intraatomic and NN couplings, more distant neighbors should be included into consideration. Clearly 3-5 shells of neighboring atoms have to be considered. Such itinerant effects would strongly affect the temperature dependence of magnetic anisotropy in these systems. We also note, that the difference between the results for $\lambda=0$ and $\lambda=1$ in pure $\text{Fe}_{16}\text{N}_{2}$ and in the doped systems allows to separate the effects of chemical bonding (hybridization) contributions to the intersite anisotropies through doping from changes created by SO coupling of the dopant. For example, the hybridization effect can be seen from the non-slope curves of atoms 11, 14 and 15 in the case of Sb and Bi doping, while the strong effect of SO coupling on the atom 10 in Pt doping can be clearly identified from the large slope in Fig. 2.

\section{Conclusion}
We proposed a method of decomposition of the total relativistic energy in solids to intra and interatomic orbital contributions.  The technique is based on a site/orbital variation of the speed of light in a particular term of relativistic electronic Hamiltonian (in our case the spin-orbit coupling). It does not require the use of traditional Green function methods and naturally allowed us to study the spatial decomposition of the total energy in precise modern band structure methods without any approximations for the shape of the potential. It also does not require any approximate treatment of the total energy including very popular analysis of its one-electron component.  This technique can be used when other relativistic interactions, such as dipole-dipole or spin-other-orbit are considered.  As an illustration of the method, we tested such decomposition in the case of the magnet CoPt, and then we analyzed the pairwise interactions in $\alpha''-\text{Fe}_{16}\text{N}_{2}$  doped by Bi, Sb, Co and Pt on the 4d site. The site decomposition revealed the most important pairwise interactions and showed different mechanisms of increasing magnetocrystalline anisotropy for the Pt and Bi/Sb dopants. We found that the anisotropic interactions in studied metallic systems are relatively long ranged with pairwise contributions often being larger than on-site ones. Theoretically considered dopants increased the magnetocrystalline anisotropy of the original $\text{Fe}_{16}\text{N}_{2}$  phase. Our studies of structural properties of these alloys predicted the possible stability of some doped systems. This created an opportunity for the experimental verification of our predictions.

\section{acknowledgments}
Majority of this work was done at Ames Laboratory and supported by the U.S. Department of Energy (DOE), Office of Science, Basic Energy Sciences, Division of Materials Science and Engineering, and included the computer time support from the National Energy Research Scientific Computing Center (NERSC) in Berkeley, CA. Ames Laboratory is operated for the DOE by Iowa State University under Contract No. DE-AC02-07CH11358. M.C.N. also acknowledges the partial support from the National Science Foundation (NSF) award DMR-1729677. Y.S. was partially supported by NSF awards EAR-1918134 and EAR-1918126.

\renewcommand{\bibnumfmt}[1]{[#1]}
\bibliographystyle{apsrev4-1}
\end{document}